\documentclass[twocolumn,showpacs,preprintnumbers,amsmath,amssymb]{revtex4}
\usepackage{tabularx,graphicx}\begin{document}
\newcommand{\beq}{\begin{equation}}
\newcommand{\eeq}{\end{equation}}
\newcommand{\beqn}{\begin{eqnarray}}
\newcommand{\eeqn}{\end{eqnarray}}
\newcommand{\bmath}{\begin{subequations}}
\newcommand{\emath}{\end{subequations}}
\title{The missing angular momentum of superconductors}
\author{J. E. Hirsch }
\address{Department of Physics, University of California, San Diego\\
La Jolla, CA 92093-0319}

\begin{abstract} 
We point out that the Meissner effect, the process by which a superconductor expels magnetic field from its interior, represents an unsolved  puzzle within the London-BCS theoretical framework used
to describe the physics of conventional superconductors, because it appears to give rise  to non-conservation of angular momentum. Possible ways to avoid this inconsistency  within the
conventional theory of superconductivity are argued to be farfetched. Consequently, we argue that unless/until a consistent explanation is put forth,
 the existence of the Meissner effect represents an anomaly that casts doubt on the    validity of the conventional framework. 
Instead, we point out that two elements of the unconventional theory of hole superconductivity, namely that the charge 
 distribution in superconductors is macroscopically inhomogeneous, and the fact that spin-orbit coupling plays an essential role, allow for a consistent explanation of the Meissner effect.

   \end{abstract}
\maketitle 
\section{introduction}
Superconductors do not allow the presence of magnetic fields in their interior\cite{tinkham}. This fundamental property  distinguishes them from
`perfect conductors' and was unexpected before it was experimentally discovered by
Meissner and Ochsenfeld in 1933\cite{meissner}. Soon thereafter the phenomenology was described by  London's equation\cite{london} that relates 
  the current density $\vec{J}$ to the magnetic vector potential $\vec{A}$\cite{gauge}:
  \beq
\vec{J}=-\frac{c}{4\pi \lambda_L^2} \vec{A}
\eeq
where  $\lambda_L$ is  the London penetration depth. Eq. (1) implies that the magnetic field $\vec{B}=\vec{\nabla}\times\vec{A}$ obeys the equation
$\nabla^2\vec{B}=(1/\lambda_L^2)\vec{B}$, and consequently that magnetic fields cannot exist in the interior of a superconductor  beyond a distance $\lambda_L$ from the surface.
The BCS microscopic theory of superconductivity gives a wavefunction for the superconducting state that predicts\cite{tinkham} the current response Eq. (1) to an applied $\vec{A}$, and for these
reasons it is generally believed that BCS-London theory accounts for the experimentally observed Meissner effect\cite{meissner2}.

However:  neither BCS theory nor London theory address the question of $how$ the system attains the final (superconducting) state where the magnetic field is excluded starting from an initial 
(normal) state where magnetic field exists in the interior\cite{lenz}. Specifically: (i)  What is the nature of the $force$ that causes the superfluid electrons near the surface to all acquire a velocity in the direction
required to screen the external magnetic field?  And: (ii)   How is the angular momentum of the electrons in the Meissner current compensated? In particular, we  argue here that the conventional framework appears to be  incompatible with
 angular momentum conservation, and hence cannot explain the Meissner effect in a consistent way. 
 
 There have been   attempts to explain the Meissner effect using classical electrodynamics\cite{edwards,bostick}, motivated by the fact that Planck's constant does not
 enter the expressions for the  Meissner current nor the London penetration depth. Furthermore,   Meissner-effect-like properties of magnetized classical plasmas have been noted\cite{witalis,vahala}. Much earlier, Heisenberg\cite{heisenberg} attempted to explain the Meissner effect using only the classical Lorentz force. 
 However, there is general consensus (with which we agree) that a purely classical theory cannot explain the Meissner effect\cite{london2,putterman,taylor}, hence that quantum mechanics plays a fundamental role. Still, we argue  that (even though this is not generally recognized) precisely $how$ quantum mechanics explains the Meissner effect is not understood.
Moreover, we argue that a consistent explanation of the Meissner effect 
 may   in fact  be beyond the  confines of the conventional theory to an extent that calls the  validity of   the entire conventional framework into question. 
   
 The question of what  is the $force$ propelling electrons to develop the Meissner current\cite{lorentz}  in the transition to the superconducting state in the presence of a magnetic field   is strangely absent in the
 literature on superconductivity, even in the early days. As an exception, we mention a paper by H. London in 1935\cite{london35} where he discusses the  motion of  the phase boundary between normal and
 superconducting phases in a magnetic field and states ``The generation of current in the part which becomes supraconductive takes place without any assistance of an electric field and is only due to
 forces which come from the decrease of the free energy caused by the phase transformation'', but does not discuss the nature of these ``forces''. Recently, Nikulov postulated a ``quantum force'' for
 superconductors so that   ``superconducting pairs are accelerated against the force of the electric field...''\cite{nikulov}. Except for these rare instances, we are not aware of any discussion of this question in the
 literature. The related question of angular momentum conservation has never been raised to our knowledge. The purpose of this paper is to call attention to these questions and propose answers to them.   

 \section{angular momentum in  the Meissner current}

We assume 
 that the orbital magnetic response currents are   carried by bare electrons of mass $m_e$ and charge $e$ with volume number density $n_s$, both in the normal and
in the superconducting state. The London
penetration depth is given by\cite{tinkham}
\beq
\frac{1
}{\lambda_L^2}=\frac{4\pi n_s e^2}{m_e c^2}
\eeq
and is of order several hundred Angstrom in a conventional type I superconductor.
Consider a long metallic cylinder with a magnetic field $\vec{B}$ pointing along its axis.
In the normal state, the Landau diamagnetic susceptibility\cite{landau}
\beq
\chi_{Landau}=-\frac{1}{3}\mu_B^2g(\epsilon_F)
\eeq
($\mu_B=e\hbar/2m_e c=$Bohr magneton, $g(\epsilon_F)=$density of states at the Fermi energy) can be interpreted as arising from Larmor orbits perpendicular to the applied
magnetic field
\beq
\chi_{Larmor}=-\frac{n_s e^2}{4 m_e c^2} a^2
\eeq
of radius  $a=1/k_F$, for a free electron density of states at the Fermi energy $g(\epsilon_F)=3n_s/2\epsilon_F$, with $\epsilon_F=\hbar^2k_F^2/2m_e$. In the perfectly diamagnetic
superconducting state, the magnetic susceptibility is
\beq
\chi_{London}=-\frac{1}{4\pi}=-\frac{n_s e^2}{4 m_e c^2} (2\lambda_L)^2,
\eeq
and is
larger than the normal state susceptibility Eq. (3)  by a factor $(2\lambda_L k_F)^2$.  

Similarly, the mechanical angular momentum density induced by a perpendicular magnetic field $\vec{B}$ on electrons in    orbits of radius $a$  in the plane perpendicular to $\vec{B}$  is
\beq
\vec{l}_e=-\frac{en_s}{2c}a^2\vec{B}
\eeq
and in the normal state the mechanical angular momentum density induced by the applied magnetic field  is 
\beq
\vec{l}_e^n=-\frac{en_s}{2c}(k_F^{-1})^2\vec{B} 
\eeq
arising from electrons in orbits of radius $k_F^{-1}$. 
In the superconducting state, the induced surface current density that suppresses the interior magnetic field has magnitude
\beq
J=|e|n_sv_s=\frac{c}{4\pi \lambda_L}B
\eeq
where $v_s=|e|\lambda_L B/m_e c$  is the velocity of the superfluid electrons near the surface. For a cylinder of radius $R$, each electron in the Meissner current carries angular momentum $m_e v_s R$, and there are
$N=2\pi R \lambda_L h n_s$ electrons in the surface layer of thickness $\lambda_L$ for a cylinder of height $h$. Hence the total electronic angular momentum per unit volume is
\beq
\vec{l}_e^s=-\frac{m_e c}{2\pi e} \vec{B}=-\frac{en_s}{2c}(2\lambda_L)^2\vec{B}
\eeq
 and again  is larger than that in the normal state by a factor   $(2\lambda_L  k_F )^2$, which is of order $10^5$ or larger for a typical type I superconductor. {\it Where did the extra angular momentum
come from?}

In the foregoing we have assumed that the mechanical angular momentum in the Meissner current is carried by bare electrons of mass $m_e$. This has been
experimentally demonstrated by measuring the angular momentum acquired by a superconducting body when a magnetic field is applied
(gyromagnetic effect)\cite{gyro1,gyro2,gyro3}. The angular momentum of the body is found to be given by Eq. (9) with opposite sign (i.e antiparallel to the applied magnetic field),  corresponding 
to the equal and opposite momentum acquired by the
positive ions. This result can be simply understood as arising from the effect of the induced Faraday electric field on electrons and ions when a magnetic field attempts to
penetrate a superconductor. The corresponding experiment for the case where   the magnetic field is $expelled$ from the superconductor has not been performed, nor has  the 
question been considered theoretically (except for  ref.\cite{lenz}).

\section{meissner effect in the conventional framework}
The conventional theory has not addressed the question of  how conservation of angular momentum is preserved in the transition to the superconducting
state in the presence of an external magnetic field. There is no obvious 'force' that will cause the
electrons near the surface to start moving all in the same direction to generate the Meissner current, and would at the same time give rise to a "reaction force"   to maintain
the total angular momentum equal to zero.  It appears to be generally assumed
that since the free energy of the superconductor is lower in the state where the magnetic field is excluded, the system
will 'find its way' through statistical fluctuations to this low energy state where the Meissner current flows. Even if one were to accept this reasoning, it does
not explain how angular momentum is conserved. Let us try to understand this question within the conventional theory.

We will not attempt to model in detail the process by which the superconductor expels magnetic field from its interior. The question has been addressed experimentally\cite{faber,faber2}
and it appears that highly irregularly shaped structures form in the transition process depending on the experimental conditions. For the purposes of this paper we are interested
in a conservation law relating the initial and final states and because of this the details of the intermediate processes are largely irrelevant.

In the process of cooling the system from above to below $T_c$, no angular momentum is transferred from the environment. Similarly, in changing the external magnetic field
from just above the critical magnetic field $H_c$ to just below $H_c$ only a tiny amount of angular momentum can be generated, which cannot account
for the difference between Eq. (7) and Eq. (9)\cite{lenz2}. The angular momentum of the electromagnetic field
\beq
\vec{L}_{field}=\frac{1}{4\pi c}\int d^3r \vec{r}\times(\vec{E}\times\vec{B})
\eeq
is zero both in the normal and in the superconducting states, since no electric field $\vec{E}$ exists after the system has reached equilibrium within the conventional theory. Furthermore we can assume that the transition occurs sufficiently slowly   that no electromagnetic momentum is carried away by radiation during the transition process.
Consequently, the difference between
the angular momenta Eq (7) and Eq. (9) has to be picked up by the ionic lattice.

There are two ways in which the ionic lattice can acquire angular momentum: through interaction with the electromagnetic field, and through direct interaction with the  electrons.
We  discuss these in turn.

  \begin{figure}
\resizebox{8.5cm}{!}{\includegraphics[width=7cm]{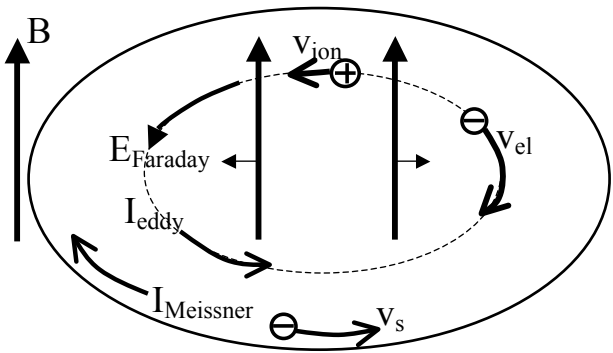}}
\caption {Charge velocities in the transition to the superconducting state in the presence of a magnetic field B pointing up, that is being expelled from the interior of the superconductor.
The electrons in the Meissner current have counterclockwise velocity $v_s$. In the interior a counterclockwise electric field $E_{Faraday}$ is generated as the magnetic field is expelled,
that imparts a transient equal and opposite angular momentum to interior electrons and ions, that is cancelled as the system reaches the final equilibrium  state. The interior electron motion is manifested
in transient eddy currents. }
\label{figure2}
\end{figure}

As the electrons develop a Meissner current, the magnetic field becomes smaller in the interior of the superconductor (magnetic field lines are pushed out). By Faraday's law, this
change in magnetic flux generates an electric field in direction such that it opposes the change in magnetic flux, as shown in Fig. 1. This electric field pushes the positive ions in the same direction as 
the electrons in the Meissner current, i.e. to acquire mechanical angular momentum $parallel$ to the electronic angular momentum in the Meissner current\cite{lenz}, so it is clear that 
this effect has the wrong sign to resolve the angular momentum question.

The Faraday field also imparts angular momentum to the $electrons$ in the interior, which is $antiparallel$ to the electronic angular momentum of the Meissner current. Since the system is charge neutral,
this electronic angular momentum is equal and opposite to the angular momentum imparted by the Faraday field to the ions. This interior motion of the electrons gives rise to eddy currents (antiparallel to the Meissner current) during the 
transition process, as indicated schematically in Fig. 1. 
Eventually these eddy currents   die out by collisions between the electrons and the ions as the system reaches equilibrium, and in the process the electronic and ionic angular momenta 
  acquired due to the Faraday field cancel out. 
  
The Faraday field also acts on   the electrons and ions within a London penetration depth of the surface, and in that region it also transfers
 equal and opposite angular momenta to the electrons and ions. However  the angular momentum transferred to the electrons is of opposite sign to the one acquired by the electrons in the Meissner current, and the angular momentum 
 transferred to the ions is of opposite sign of what is needed to conserve angular momentum.
We conclude from these considerations that interaction of electrons and ions with the electromagnetic field cannot solve the angular momentum question, and consider next
  possible direct angular momentum transfers between electrons and ions.

The conduction electrons interact with the periodic ionic lattice and its static (due to impurities) and dynamic (due to lattice vibrations) deviations from periodicity.  
 Let us assume first there are no impurities. The effective electron-electron interaction
resulting from the electron-phonon interaction will not change the center of mass momentum of the interacting electrons nor transfer momentum between electrons and ions.
The periodic ionic lattice can be regarded as a $classical$ system of positive charges, which is initially at rest. 
Hence we argue that the process by which the ions acquire angular momentum should be readily understandable
through classical electromagnetism. Electrons interact with the ions at rest
through electrostatic attraction. The time derivative of the angular momentum of a given ion $\alpha$ with charge $Z|e|$  is simply the sum of the torques due to the negative electrons:
\beq
\frac{d\vec{L}_\alpha}{dt}=\vec{R}_\alpha\times\vec{F}_\alpha=\vec{R}_\alpha\times\sum_i\frac{Ze^2}{|\vec{R}_\alpha-\vec{r}_i|^3}\vec{r}_i
\eeq
where $\vec{R}_\alpha$ and $\vec{r}_i$ are the position vectors of the $\alpha$-th ion and the $i$-th electron. Within the conventional theory of superconductivity
 the spatial distribution of electrons is homogeneous 
at all times, which implies that the sum in Eq. (11) is zero. Hence neither the angular momentum of a given ion nor the total angular momentum of the ions 
\beq
\vec{L}_{ions}=\sum_\alpha \vec{L}_\alpha
\eeq
can change through the electrostatic interaction with the electrons. If the ions never acquire motion, no $magnetic$ Lorentz force can act on them either.

However, the above argument holds rigorously only in the absence of disorder, and one  may argue that while in such cases
the superconducting state cannot be reached, for any non-zero disorder   electrons would be able to transfer the required angular momentum
to the ions as the system develops the Meissner current\cite{fogler}.  Let us examine such a possible scenario to assess its feasibility.

The superfluid electrons are insensitive to non-magnetic disorder\cite{anderson} and hence will not transfer momentum to the ions through interaction with impurities. However,
a system of 'normal' electrons  moving with total momentum $\vec{P}$ will certainly transmit  momentum to   ions at rest  in the presence of impurity scattering. 
 Consider a metal at zero temperature in an external magnetic field that is lowered from right above to right below the critical field. 
 The electrons can extract angular momentum from the electromagnetic field {\it only if there is a radial flow of charge}\cite{lenz} (see next section). However no radial charge flow is
 predicted in the conventional theory. Hence, to explain the Meissner effect and conserve angular momentum we need to assume that through some quantum-mechanical
process  some electrons will acquire angular momentum in the direction of the Meissner current, while other 'normal' electrons will acquire equal and opposite angular momentum which they then
 transfer to the ionic lattice. Furthermore, because we are at zero temperature, no ``normal'' conduction  electrons can remain at the end of the process.

The maximum momentum that an electron with momentum $\vec{p}$ can transfer to a much more massive ion in a head-on collision is $2\vec{p}$, whence the electron acquires momentum
$-\vec{p}$. A possible 'cartoon' scenario might be that for each conduction electron (or half-a-Cooper pair) that condenses into the superconducting state and acquires velocity $\vec{v}_s$ there is another
electron that remains normal and acquires  (through the same quantum-mechanical process that imparted momentum to the electron becoming superfluid)
equal and opposite velocity and momentum. Then, the 'normal' electron could
'bounce off' an ionic impurity, reverse its momentum and subsequently condense into the superfluid state. In this way, the ions would acquire angular momentum density
Eq. (9) with opposite sign, satisfying angular momentum conservation.

However,  it is statistically impossible that each normal electron would exactly reverse its momentum in collisions with the ions. Rather, a system of normal electrons with total momentum
$\vec{P}$ that undergoes random collisions with ions will eventually relax and come to rest relative to the ions, transmitting  its momentum $\vec{P}$ (rather than $2\vec{P}$) to the
much more massive ions. So to satisfy angular momentum conservation we need to assume that the electrons that initially remain normal acquire through a quantum-mechanical process
{\it the full} angular momentum $-\vec{l}_e^s$ (Eq. (9)).  This requires that both the electrons that initially become superfluid and those that remain normal acquire speeds on average larger than $v_s$
($2v_s$ each on average  if their number is equal). 
Then, one could imagine that as the normal electrons lose their momenta in scattering off the ions they will condense into the superconducting state and share in the motion of the
superfluid,  and at the end of the process all 
conduction electrons will be superfluid with angular momentum Eq. (9), and the ions will have acquired equal and opposite angular momentum.

One can devise other more elaborate variants of these scenarios.  However,   they all  require that    in the process of condensation   the electrons that become superfluid  
first acquire speeds on average   larger than
$v_s$. We argue that such
scenarios are farfetched and are  certainly not described by the conventional theory in its current form:  there is no mechanism in the conventional
theory for the condensing superfluid electrons to acquire average speed larger than $v_s$, since the speed $v_s$ is constrained by a quantum condition on the phase of the superfluid wave function.  
Thus,  we argue that within the conventional theory of superconductivity there exists an unaccounted
angular momentum
\beq
\vec{L}_{missing}=V(\vec{l}_e^s-\vec{l}_e^n)=V\frac{|e|n_s}{2c}k_F^{-2}\vec{B}((\frac{2\lambda_L}{k_F^{-1}})^2-1)
\eeq
($V$=sample volume)
 when a metal enters the superconducting state in the presence of an external magnetic field $\vec{B}$. Consequently, that (at least in its present form)  {\it the conventional theory
is internally inconsistent}.

\section{meissner effect and $r=2\lambda_L$ orbits}

It is a remarkable fact that within the conventional theory of superconductivity it has not been recognized  that
electronic orbits of radius $2\lambda_L$ play a key role.  This was proposed in ref.\cite{sm} and shown to lead
to the predicted "spin Meissner effect".

The simplest argument leading to $2\lambda_L$ orbits is the following. For a cylinder of radius $R$ and height $h$, and Meissner
current residing in a surface layer of thickness $\lambda_L$ with $n_s$ carriers per unit volume moving with speed $v_s$, the
total mechanical angular momentum carried by the surface current is
\bmath
\beq
L_{Meissner}=[n_s 2\pi R \lambda_L h]\times[m_e v_s R]
\eeq
where the first factor is the number of electrons in the surface layer, and the second factor the angular momentum of each electron
in the surface layer. By simply changing the order of the factors this can be rewritten as
\beq
L_{Meissner}=[n_s\pi R^2 h]\times[m_e v_s  (2\lambda_L)]
\eeq
\emath
where the second factor in square brackets  is the angular momentum of
an electron in an orbit of radius $2\lambda_L$, and the  first factor   is the total number of such orbits
(i.e. the total number of electrons) {\it in the bulk}.

Orbits of radius $2\lambda_L$ also follow directly from the fact that when a magnetic field is applied to a superconductor, an equal and opposite
magnetic field is generated in the interior. Let us go
through the simple argument. The relation between orbital magnetic moment $\vec{\mu}$ 
 and orbital angular momentum $\vec{l}_e$ for an electron of charge $e$ and
mass $m_e$ is
\beq
\vec{\mu}=\frac{e}{2m_e c}\vec{l}_e
\eeq
In an orbit of radius $a$ with speed $v$, the orbital angular momentum is $l_e=m_e v a$ and the magnetic moment is
\beq
\mu=\frac{ev}{2c}a
\eeq
Application of an external magnetic field generates a Faraday field, satisfying
\beq
\oint \vec{E}\cdot d\vec{l}=-\frac{1}{c}\frac{\partial}{\partial t}\int \vec{B}\cdot \hat{n} dS
\eeq
and for an orbit of radius $a$
\beq
E=\frac{a}{2c} \frac{\partial B}{\partial t}
\eeq
so the change in speed for an electron in such an orbit is
\bmath
\beq
\frac{dv}{dt}=\frac{e}{m_e}E=\frac{ea}{2m_e c} \frac{\partial B}{\partial t}
\eeq
\beq
\Delta v=\frac{ea}{2m_ec}B
\eeq
\emath
Note that to lowest order the radius of the orbit does not change as the magnetic field is applied, because the magnetic
Lorentz force precisely cancels the increased centripetal acceleration resulting from the change in speed:
\beq
\Delta(m_e\frac{v^2}{a})=2m_e \frac{v\Delta v}{a}=e\frac{v}{c} B
\eeq
for $\Delta v$ given by Eq. (19b).

Consequently the induced magnetic moment per electron $\Delta \mu$ and the induced magnetization per unit volume $M$ are
\bmath
\beq
\Delta \mu=\frac{ea}{2c}\Delta v=\frac{e^2a^2}{4m_e c^2}B
\eeq
\beq
M=n_s\Delta \mu =\frac{n_s e^2 a^2}{4m_e c^2}B
\eeq
\emath
For a long cylinder, the magnetic field in the interior generated by a uniform magnetization $M$ is $B_{ind}=4\pi M$.
Hence to completely suppress the applied field $B$ we require $ M=B/4\pi$, hence
\beq
\frac{B}{4\pi}=n_s\Delta \mu =\frac{n_s e^2 a^2}{4m_e c^2}B
\eeq
so the required radius of the orbit is
 \beq
a=\sqrt{\frac{m_e c^2}{\pi n_s e^2}}
\eeq
or, using Eq. (2)
\beq
a=2\lambda_L.
\eeq

Equivalently, the fact that superconducting electrons reside in orbits of radius $2\lambda_L$ can also be deduced from an
energetic argument. In changing the applied magnetic field from $B$ to $B+\Delta B$, the electron in an orbit of radius $a$
changes its energy by
\beq
\Delta \epsilon=B\Delta \mu=\frac{e^2 a^2}{4m_e c^2} B\Delta B
\eeq
Integrating   from $0$ to $B$ we obtain for the increase in energy per unit volume, for $n_s$ electrons per unit volume each residing
in an orbit of radius $a$
\beq
u\equiv n_s\Delta \epsilon=\frac{n_s e^2 a^2}{8m_e c^2} B^2
\eeq
The system will remain superconducting until this energy cost equals the superconducting condensation energy per unit volume, 
$H_c^2/8\pi$, with $H_c$ the thermodynamic critical field\cite{tinkham}. This  will of course occur when $B=H_c$, hence
\beq
u=\frac{n_s e^2 a^2}{8m_e c^2} H_c^2=\frac{H_c^2}{8\pi}
\eeq
leading again to Eq. (23) for the radius of the orbits, and hence to $a=2\lambda_L$.

The arguments spelled out in detail here merely restate the fact that Eqs. (5) and (9) can be interpreted as resulting from electrons occupying Larmor orbits of radius $r=2\lambda_L$.

\section{meissner effect in the theory of hole superconductivity}
Besides the importance of $2\lambda_L$ orbits, 
two other elements of the theory of hole superconductivity\cite{hole1,holetheory},
namely  the predicted existence of charge inhomogeneity\cite{chargeexp} and  the essential role of spin-orbit coupling\cite{sm}, 
play a key role in understanding  the Meissner effect. 
Charge inhomogeneity is accompanied by the presence of an internal electric field,
and hence allows for some angular momentum to be carried by the electromagnetic field (Eq. (10)); the spin-orbit interaction is a   $velocity-dependent$ electron-ion interaction that
allows for transmission of angular momentum from the electrons to the ions even in the absence of disorder. It should be pointed out that
 neither of these elements was introduced in the
theory $in$ $order$ to account for the Meissner effect\cite{holetheory}. 

 Qualitatively, it is easy to see that $radial$ motion of charge is likely to play an essential role in the Meissner effect\cite{lorentz}. A radially outgoing electron in a magnetic field $\vec{B}$ acquires 
 through the action of the magnetic Lorentz force an azimuthal 
 velocity in direction $- \hat{r}\times\hat{B}$, which is the azimuthal direction of the electrons in the Meissner current. The theory of hole superconductivity predicts that 
 negative charge is expelled from the interior towards the surface when a metal makes a transition to the superconducting state, whether or not an external magnetic field is present\cite{chargeexp}.
 
Eqs. (4) and (5) for the diamagnetic susceptibility in the normal and superconducting state
indicate  that the transition to superconductivity can be understood as an $expansion$ of the radius of the electronic orbit from a microscopic $a= k_F^{-1}$ to a mesoscopic
$2\lambda_L$\cite{sm}. This interpretation is corroborated by Eqs. (7) and (9): the angular momentum of the electrons in the Meissner current in the surface layer arises
 from mesoscopic orbits of radius $2\lambda_L$ for each electron in the bulk that expanded from a microscopic radius $a=k_F^{-1}$ in the normal state (Eq. (7)). 
As the expanding electronic orbit cuts through magnetic field lines the electron acquires angular momentum due to the Lorentz force acting on it, satisfying
\beq
l_{final}-l_{initial}=-n_s \frac{e}{2\pi c} (\phi_{final}-\phi_{initial})
\eeq
where $\phi$ is the magnetic flux enclosed in the orbit. Eq. (28) exactly accounts for the difference between the angular momenta Eqs. (7) and (9) for initial radius $a=k_F^{-1}$ and final radius $2\lambda_L$, and provides a 'dynamical' explanation of the Meissner effect\cite{sm} (i.e., it explains the origin of the force that causes the electrons to move in the
direction required for the Meissner current).
However we still need to understand how this extra electronic angular momentum  is compensated.

  \begin{figure}
\resizebox{8.5cm}{!}{\includegraphics[width=7cm]{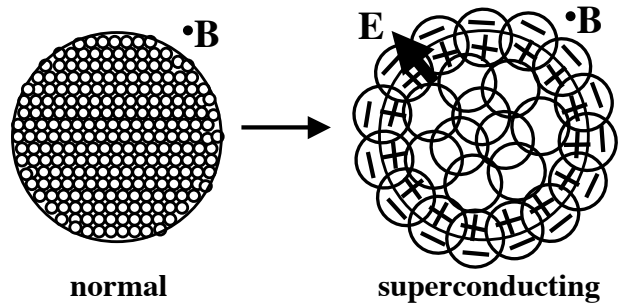}}
\caption{In the normal state, electrons in a magnetic field traverse microscopic orbits of radius
$a= k_F^{-1}$ (left side). When the system goes superconducting, the orbits expand to
radius $2\lambda_L$ (right side). Assuming the centers of the orbits don't move, negative charge
spills out and a surface 
'double layer' of charge of thickness $\sim 2\lambda_L$ is created with outward pointing electric field. }
\label{figure2}
\end{figure} 

Assume every electron in the cylindrical sample undergoes such an orbit expansion. The electrons within a distance $2\lambda_L$ of the surface will have their orbits ``spill out'' beyond the surface of the superconductor,
leaving behind a positive surface layer of charge  density  $\sigma=|e|n_s\lambda_L$,  which will give rise to a ``double layer'' with an electric field
\beq
E=4\pi |e| n_s \lambda_L=\frac{m_e c^2}{e\lambda_L}
\eeq
pointing radially outward, as shown schematically in Fig. 2. The electric field can be assumed to be uniform over a thickness $\lambda_L$, and it gives rise to an angular momentum of the electromagnetic field (Eq. (10))
\beq
\vec{L}_{field}=\frac{2\pi e n_s}{c}\lambda_L^2 R^2 h \vec{b}=V\frac{en_s}{2c}(2\lambda_L)^2\vec{B}
\eeq
which is equal and opposite to the angular momentum of the Meissner current Eq. (9). Thus, (neglecting the small angular momentum in the normal state) in this scenario the angular momentum in the electromagnetic field
 accounts for the ``missing'' angular momentum Eq. (13), and the angular momentum puzzle is resolved. In other words, the 'reaction' to the angular momentum
imparted by the electromagnetic field to the expanding electron orbit is stored as equal and opposite angular momentum  in the electromagnetic field.\cite{feynman}

Unfortunately, this is not a realistic scenario. The electric energy density in the assumed double layer is an enormous $E^2/8\pi=n_s m_e c^2/2$, and the electric energy density per unit volume in the entire sample is
\beq
u=n_s m_e c^2 \frac{\lambda_L}{R}
\eeq
which is much larger than the superconducting condensation energy density  even for a sample of $R\sim 1cm$. The electric field in the double layer Eq. (29) is of order $10^{11}V/cm$ which is clearly unsustainable. It is clear that the interaction of electrons with the positive ionic lattice will $prevent$ the electrons from spilling out a distance $2\lambda_L$ as depicted
in Fig. 2. What is not yet clear is how the ions, in the process of  preventing the electrons from spilling out to the extent shown in Fig. 2,    will acquire compensating angular momentum in the required direction.

In Ref. \cite{lenz} we explored a related scenario, using the fact that the theory of hole superconductivity
predicts that a  positive charge density $\rho_0$  exists uniformly distributed in the interior of the superconductor and a negative charge density $\rho_-$ in a surface
layer of thickness $\lambda_L$\cite{chargeexp}. To account for a suppression of the internal magnetic field to a fraction $y$ of its original value and compensating the electronic angular momentum with momentum in the
electromagnetic field requires an electric field near the surface\cite{lenz}
\beq
E_m=\frac{4 m_e c^2}{eR} \frac{1-y}{y}
\eeq
hence for example for a $99\%$ suppression ($y=0.01$) with $R=1cm$, $E_m=2\times 10^8 V/cm$. While this electric field is three orders of magnitude smaller than Eq. (29) it  is still too large, and in addition this scenario
cannot account for a full Meissner effect, since $E_m$ diverges as $y\rightarrow 0$ (Eq. (32)).
We conclude from these considerations that it is impossible to explain the Meissner effect in superconductors without a mechanism that allows the ions to acquire angular momentum in direction opposite to the
applied magnetic field through interaction with the electrons.

 \begin{figure}
\resizebox{5.5cm}{!}{\includegraphics[width=7cm]{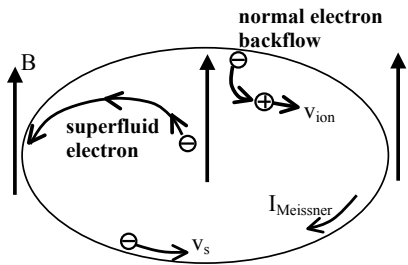}}
\caption {Superfluid electrons flow from the interior towards the surface and are deflected to the left by the magnetic field pointing up. Normal electrons backflow from the surface towards
the interior and are deflected to the right by the magnetic field. The momentum in this normal current is transferred to the ions by collisions with impurities.   }
\label{figure2}
\end{figure} 

One possible way for ions to acquire angular momentum is depicted in Fig. 3. Suppose that when superfluid electrons are expelled towards the surface there is a radial backflow of 'normal'
electrons attempting to maintain charge neutrality. The normal electrons will be deflected by the Lorentz force in opposite direction to the superfluid electrons, as shown in Fig. 3.
In the presence of disorder, these normal electrons will scatter off the ions and transmit their momenta to the ions, which will thus acquire angular momentum in direction opposite to the
Meissner current. To achieve a full Meissner effect, it is necessary that a fraction $\sim \lambda_L/R$  electrons in the surface layer of thickness $\lambda_L$ flow there from the
interior\cite{ionizing}, and the same amount has to flow in from the surface layer and transmit their momenta to the ions. 

While somewhat less farfetched than the scenario described earlier  in the conventional framework (at least this scenario provides an origin for the   azimuthal forces giving rise
to the Meissner current and ionic countercurrent), this scenario also requires a division of  electrons into `normal' and
`superfluid'. While it may contribute at finite temperatures, it cannot play a role as the temperature approaches zero in the process where the
external magnetic field is changed from just above to just below
the critical field. Instead, we argue in what follows that the spin-orbit interaction plays a crucial role in the Meissner effect.

\section{role of the spin-orbit interaction}

The spin-orbit interaction offers a natural solution to the puzzle. In the spin-Meissner effect scenario proposed in Ref.\cite{sm}, as the electron orbit radius expands
from $ k_F^{-1}$ to $2\lambda_L$, a torque $\vec{\tau}_{ie}$ is exerted by the positive ionic charge on the equivalent electric dipole $\vec{p}$\cite{dipole} resulting from the moving
electron magnetic moment $\vec{\mu}=e\hbar/(2m_e c)\vec{\sigma}$
\bmath
\beq
\vec{p}=\frac{\vec{v}}{c}\times\vec{\mu}
\eeq
\beq
\vec{\tau}_{ie}=\vec{p}\times\vec{E}_i=(\frac{\vec{v}}{c}\times\vec{\mu})\times\vec{E}_i
\eeq
\emath
where $\vec{E}_i$ is the radial electric field generated by the positive ionic charge density $|e|n_s$
\beq
\vec{E}_i=2\pi |e|n_s \vec{r}
\eeq
This torque causes electrons of opposite spin to acquire azimuthal velocities in opposite directions, giving rise to a spontaneous spin current (spin Meissner effect)\cite{sm}.
By Newton's third law, the torque exerted by the ions on the electrons is necessarily accompanied by an equal and opposite torque exerted by
the electrons on the ions:
\beq
\vec{\tau}_{ei}=-(\frac{\vec{v}}{c}\times\vec{\mu})\times\vec{E}_i=-\vec{p}\times\vec{E}_i .
\eeq

In the absence of external magnetic field, spin up and spin down electrons acquire opposite angular momenta, and exert equal and opposite torques on the ions, hence the net
angular momentum  transferred to the ionic lattice is zero. The resulting azimuthal motion of the electrons (Fig. 4) can be understood as resulting from the action of an effective `spin-orbit' magnetic
field\cite{sm}
\beq
\vec{B}_\sigma=2\pi n_s\vec{\mu}\equiv -B_{s.o.}\vec{\sigma}
\eeq
of magnitude $B_{s.o.}$ pointing $antiparallel$ to the electron spin (parallel to its magnetic moment). Expansion of the electron orbit to radius $2\lambda_L$ results in an azimuthal velocity of
magnitude\cite{sm}
\beq
v_\sigma^0=\frac{|e|\lambda_L}{m_e c}B_{s.o.}=\frac{\hbar}{4m_e\lambda_L}
\eeq
with opposite spin electrons orbiting in opposite directions. In the presence of an external magnetic field $\vec{B}$, the effective magnetic field acting on the electrons has
magnitude$(B_{s.o.}\pm B)$, with the $+$ sign corresponding to electrons with spin antiparallel to $\vec{B}$. The resulting azimuthal velocities are
\beq
v_\sigma=\frac{|e|\lambda_L}{m_e c}(B_{s.o.}\pm B).
\eeq
Because the speed acquired by opposite spin electrons is different, the net torque exerted by electrons on the ions Eq. (35) no longer vanishes.
The speed acquired by electrons with magnetic moment $parallel$ to the magnetic field is larger, and consequently the net torque exerted by electrons
on ions points $antiparallel$ to the applied magnetic field. Thus, the lattice acquires  angular momentum in direction opposite to the net angular momentum
acquired by the electrons.

 \begin{figure}
\resizebox{5.5cm}{!}{\includegraphics[width=7cm]{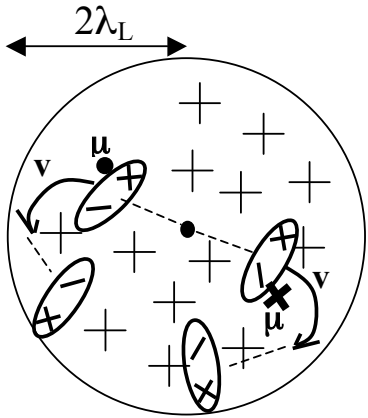}}
\caption {Up and down magnetic moments
get deflected in opposite direction due to the torque exerted by the
radially pointing electric field from the positive charge distribution. The figure  shows the equivalent electric dipoles Eq. (33a) as they move
out and after they acquired the azimuthal velocity.   }
\label{figure2}
\end{figure}

Fig. 5 illustrates in more detail how the net torque on the ions arises. Electrons with magnetic moment pointing out of (into) the paper move outward a distance $\Delta r$
along trajectories labeled 1 and 3 respectively. In the process they acquire a perpendicular impulse
\beq
\Delta I=\int F dt=\frac{e}{c}\Delta r (B_{s.o.}\pm B)
\eeq
where the + (-) sign applies to trajectory 1 (3). This impulse causes deflection in the perpendicular direction, trajectories 2 and 4, with larger speed for the electron
along trajectory 2, resulting in a larger effective dipole moment $\vec{p}$ (Eq. (33a)). The resulting torque exerted on the ions Eq. (35) is larger in magnitude for 
the electron moving along the path 1-2 (and pointing into the paper) than for the electron along the path 3-4, that exerts a smaller torque on the ions in direction
pointing out of the paper.

 \begin{figure}
\resizebox{5.5cm}{!}{\includegraphics[width=7cm]{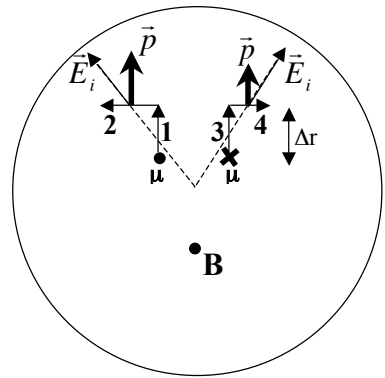}}
\caption {Electron with magnetic moment out of (into) the paper moves along paths 1 and 2 (3 and 4). The impulse in the transverse direction (Eq. (39)) acquired by the electron moving
along 1 is larger than for the electron moving along 2 for magnetic field pointing out of the paper. The resulting effective dipole moment $\vec{p}$  (Eq. (33a))  is larger for electron moving along
path 2 than it is for electron moving along path 4 as indicated schematically by the length of the vertical arrows. Consequently the torque Eq. (35) exerted by the 
electron on the ions along path 2 is larger in magnitude than that exerted by the electron  along path 4. }
\label{figure2}
\end{figure} 

 \begin{figure}
\resizebox{8.5cm}{!}{\includegraphics[width=7cm]{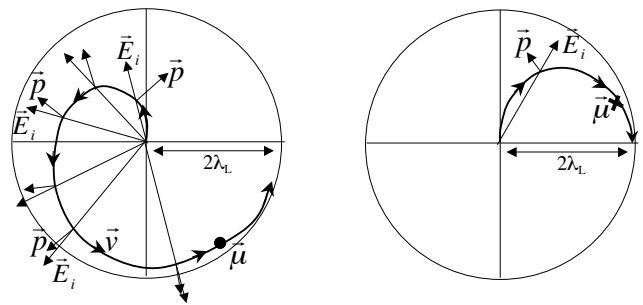}}
\caption { Simplified schematic trajectories and associated electric dipole moments $\vec{p}$ for electrons with magnetic moment out of the paper (left picture) and
into the paper (right picture) as the wavefunction expands to radius $2\lambda_L$. The speed of the electron on the right picture is smaller resulting
in smaller values of $\vec{p}$ along the trajectory and smaller torque exerted on the ions.  }
\label{figure2}
\end{figure} 

Figure 6 shows simplified schematic  trajectories of electrons of opposite spin as their wavefunction expands to radius $2\lambda_L$. For the electron
with magnetic moment pointing out of the paper, the velocity acquired is larger and so is the resulting effective electric dipole moment, resulting in a larger torque exerted
on the ions over a longer trajectory. Once the motion becomes azimuthal the effective electric dipole moment is parallel to the ionic electric field and the torque vanishes.

 \begin{figure}
\resizebox{6.5cm}{!}{\includegraphics[width=7cm]{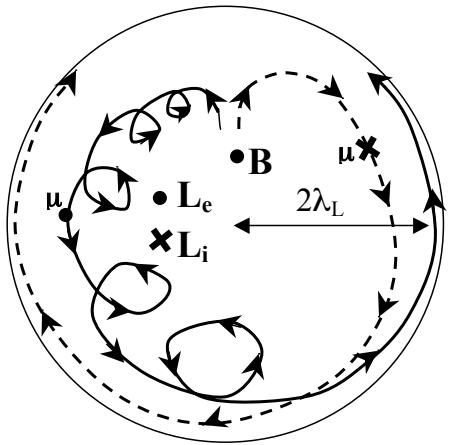}}
\caption { Schematic trajectories   for an electron with magnetic moment out of the paper (full line, left side of the picture) and
into the paper (dashed line, right side of the picture) as their wavefunction expands to radius $2\lambda_L$. The electron depicted by the full line is subject
to a larger effective magnetic field than the electron depicted by the dashed line. As it traverses its looped trajectory it gains speed and angular momentum
pointing out of the paper, and imparts to the ions a compensating angular momentum pointing into the paper. The dashed line electron is subject to a 
smaller magnetic field and its trajectory doesn't loop.}
\label{figure2}
\end{figure}

However, the actual motion, to the extent that it can be described classically, is likely to be more complicated than depicted in Fig. 6. When electron orbits expand from radius
$k_F^{-1}$ to radius $2\lambda_L$ their center will not stay fixed, since that would result in the expulsion of an enormous amount of negative charge as depicted in Fig. 2. 
The theory of hole superconductivity predicts that  the  excess negative charge density $\rho_-$ in the surface layer of thickness $\lambda_L$ that arises from
expansion of the orbits is given by\cite{unpub}
\beq
\rho_-=\frac{n_s\mu_B}{2\lambda_L}=en_s(\frac{\lambda_c}{8\pi\lambda_L})
\eeq
with $\lambda_c=h/m_e c$ the Compton wavelength of the electron. $\rho_-$ is a small fraction of the total superfluid density $en_s$ ($\sim 10^{-6}$). 
This implies that the centers of the orbits are {\it pulled inward} by the ions as the orbits expand.

The Larmor radius for spin down  ($\sigma=-1$) and spin up  ($\sigma=+1$) electrons in the
presence of magnetic and spin-orbit fields is  
 \beq
r_{\sigma}=\frac{m_e c}{e(B_{s.o.}-\sigma B)} v_\perp
\eeq
 As the orbit expansion starts, up and down spin electrons start moving outward with the same acceleration and acquire similar speeds $v_\perp$. The deflecting force is larger for the
downspin electron (magnetic moment pointing up), resulting in  a smaller Larmor radius (Eq. (41) with $\sigma=-1$), as depicted in Fig. 7. As the downspin  electron loops in
counterclockwise direction ($\vec{L}_e$ out of the page) it gains increasing azimuthal speed and it imparts clockwise angular momentum
to the ions ($\vec{L}_i$ into  the page) as shown in Fig. 7.

In the end, the angular momentum in the Meissner current is compensated partly by angular momentum in the electromagnetic field and partly by
angular momentum acquired by the ions. However, the latter is much larger than the former. The angular momentum acquired by an electron near the surface 
moving an outward distance $\lambda_L$ in the magnetic field $\vec{B}$ is 
\beq
\vec{l}_{electron}=-\frac{e}{c}R\lambda_L\vec{B}
\eeq
since the change in flux enclosed by the orbit is $\Delta \phi=2\pi R\lambda_L B$. The number of electrons acquiring this angular momentum from the electromagnetic field
for a cylinder of radius $R$ and height $h$ is $2\pi R\lambda_L h \rho_- /e$, resulting in an angular momentum density from the expelled charge given by
\beq
\vec{l}^s_{e,expelled}=-\frac{\rho_-}{2c}(2\lambda_L)^2\vec{B}
\eeq
That this coincides with the angular momentum per unit volume residing in the electromagnetic field can be seen from Eq. (10), with $\vec{l}_{field}\equiv \vec{L}_{field}/(\pi R^2 h)$:
\beq
\vec{l}_{field}=\frac{1}{2\pi c}\lambda_L E_m \vec{B}=\frac{\rho_-}{2c}(2\lambda_L)^2\vec{B}
\eeq
where $E_m$ is the (average) electric field in the surface layer of thickness $\lambda_L$, and is given by $E_m=-4\pi \lambda_L\rho_-$ for charge neutrality. Hence the angular
momentum acquired by the ions is 
\beq
\vec{l}_{ions}=\frac{en_s}{2c}(2\lambda_L)^2\vec{B}
\eeq
and the total angular momentum in the Meissner current is
\beq
\vec{l}_e^s=-\frac{1}{2c}(en_s+\rho_-)(2\lambda_L)^2\vec{B}
\eeq
compensated by $\vec{l}_{field}+\vec{l}_{ions}$. Note that the fraction of angular momentum carried by the electromagnetic field is only
 \beq
\frac{l_{field}}{l_{e}^s}\sim \frac{\rho_-}{en_s}=\frac{\lambda_c}{8\pi\lambda_L}\sim 10^{-6}
\eeq
so that $99.9999\%$ of the electronic angular momentum is in fact compensated by ionic angular momentum acquired through the spin-orbit interaction.
Figure 8 depicts schematically a superconductor in an applied magnetic field.

 \begin{figure}
\resizebox{6.5cm}{!}{\includegraphics[width=7cm]{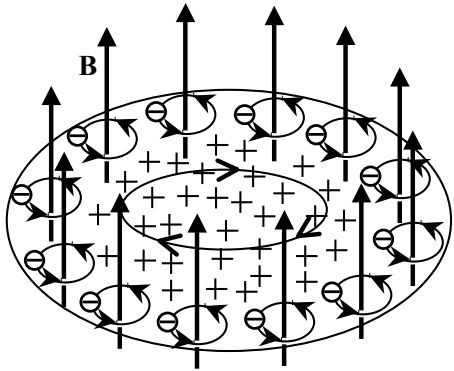}}
\caption {Schematic depiction of a superconductor in an applied magnetic field. There is excess negative charge density $\rho_-$ (Eq. (40))  in the surface layer
of thickness $\sim \lambda_L$, where the magnetic
field penetrates. Electrons reside in orbits of radius $2\lambda_L$ throughout the bulk, but only those near the surface are shown. The body as a whole rotates in clockwise direction
as indicated by the arrows in the inner circle. Electrons of spin 
down rotate in counterclockwise direction (indicated in the figure), electrons of spin up rotate in clockwise direction at slower speed\cite{sm}. The total angular
momentum is zero.}
\label{figure2}
\end{figure}

\section{Discussion}
In this paper  we have pointed out that the conventional theory of superconductivity appears to be incompatible with angular momentum conservation. It requires the
electrons near the surface to acquire a net angular momentum to generate the current to cancel  the magnetic field in the interior, but does not provide a mechanism by which this angular momentum
would be compensated. The electromagnetic field cannot carry angular momentum in the conventional theory because of the assumption that no electric field exists in the
interior of superconductors. No mechanism is provided in the conventional theory to impart the ionic lattice with angular momentum equal and opposite to the electronic angular momentum of the 
Meissner current. We have discussed possible scenarios within the conventional theory to resolve the puzzle and argued that they are farfetched.
Furthermore the conventional theory does not explain how the electrons acquire the velocity of the Meissner current in the first place.

In agreement with the general consensus, we believe that quantum mechanics is essential to understand the Meissner effect. However, 
there is no need to invoke special quantum-mechanical principles  applicable exclusively to superconductors   to explain it\cite{nikulov,berger}. Instead, we argued here
 that the underlying quantum physics responsible for  the 
Meissner effect relies on two well-known physical effects that do  $not$ play a role in the conventional theory of superconductivity, but are essential ingredients of the
theory of hole superconductivity: 
(i) The fact that a quantum particle confined to a small dimension has a high kinetic energy and exerts 'quantum pressure' to expand its
wavefunction to lower its kinetic energy;  expansion involves radial outgoing motion, and radial motion in the presence of a perpendicular magnetic field
generates an azimuthal current through the magnetic Lorentz force.  (ii) The effect of the spin-orbit interaction (in conjunction with the orbit expansion), which gives rise to a velocity-dependent interaction between 
electrons and ions that allows interchange of angular momentum between electrons and the lattice. 
Note that for  'quantum pressure' to play an important role requires that
the electronic wavefunction in the normal state is confined to small dimensions, hence an almost full band, hence hole carriers in the normal state, as required in the
theory of hole superconductivity.

The theory of hole superconductivity provides a simple and intuitive explanation for how electrons develop the Meissner current and for how angular momentum
is conserved: expansion of the electronic wavefunction, from a microscopic dimension to a mesoscopic orbit of radius
$2\lambda_L$, gives rise to outflow of 
negative charge from the interior towards the surface as the system goes superconducting, and explains dynamically how the angular momentum of the Meissner current is generated,
without the need to invoke mysterious 'quantum forces'\cite{nikulov,berger} nor statistical fluctuations. Such physics arises also in classical plasmas, where it is known as
'Alfven's theorem'\cite{alfven}: in a perfectly conducting fluid, magnetic field lines move with the fluid.
Furthermore, the outward charge flow gives rise to a spin current arising from  the spin-orbit torque exerted by the positive ions on the moving magnetic moments. We have earlier 
proposed this effect as a universal origin for the anomalous  Hall effect in ferromagnets\cite{fm} and for the spin Hall effect\cite{sh}. As pointed out here, 
Newton's law requires an equal and opposite torque
exerted by the electrons on the ions, which has opposite sign for up and down electrons, giving zero net torque in the absence of a magnetic field. In the presence of a magnetic field however, the torque exerted by  opposite spin electrons on the ions does not cancel and gives rise to a net angular momentum transfer from the electrons to the ions.

Thus, using general physical principles  and without invoking either statistical fluctuations or disorder effects, 
we are able to explain both the origin of the force giving rise to the
electronic Meissner current in superconductors,  as well as how  the angular momentum in the Meissner current is compensated, within the framework of the
theory of hole superconductivity. Part of the angular momentum  of the Meissner current is compensated by angular momentum
in the electromagnetic field, because (unlike in the conventional theory)
an electric field does exist in the interior of a superconductor within our 
 theory. The rest is compensated by angular momentum acquired by
the ionic lattice through the spin-orbit interaction that plays an essential role in our theory (and plays no role in the conventional theory).

In summary, we propose that the Meissner effect represents an 'anomaly' within the conventional theory of superconductivity: an observation that cannot be explained within
the conventional framework and is of sufficient significance to call
 the entire framework into question\cite{kuhn}. The reader may argue that the Meissner effect was discovered 75 years ago, and   it is not generally regarded to be an
  anomaly. We argue that this illustrates the phenomenon of 'retrorecognition' described by Lightman and Gingerich\cite{lg}: anomalies are often recognized as such only $after$
an explanation of them is found in  a  new theoretical framework. Before that time, according to Lightman and Gingerich
"an anomalous fact may be unquestioned or accepted as a given in the old paradigm...
not widely regarded as important or legitimized until a good explanation is at hand in a new paradigm...scientists may be so resistant to
replacing their current paradigm that they cannot acknowledge certain facts as anomalous...If unexplained facts can be glossed over or reduced in importance or simply 
accepted as givens, the possible inadequacy of the current theory does not have to be confronted"\cite{lg}.

Following that same pattern, it was only $after$ we  found  that superconductors expel negative charge from their
interior\cite{exp} according to the theory of hole superconductivity\cite{hole1}
 that we recognized   that the Lorentz force acting on the radially expelled charge provides a natural dynamical origin for  the Meissner effect\cite{lorentz}, while no comparable
intuitive explanation is  provided by
the conventional theory. Nevertheless, the  $magnitude$ of electronic angular momentum could not be
explained in the absence of a mechanism to transfer angular momentum to the ionic lattice\cite{lenz}. 
Only $after$ the essential
role of the spin-orbit interaction was recognized\cite{atom,sm} does  it become clear, as discussed in this paper,
 that this interaction provides a natural way for the ions to acquire the angular momentum needed\  to explain
the Meissner effect quantitatively.

\acknowledgements
The author is grateful to M. Fogler, H. Suhl, G. Webb, I. Schuller, C. Wu, T. O'Neil, K. Intrilligator,  and C. Surko for stimulating discussions.


\begin{references}
 \bibitem{tinkham} M. Tinkham, ``Introduction to Superconductivity'', McGraw-Hill, New York, 1996.
 \bibitem{meissner} W. Meissner and R. Ochsenfeld, Naturwissenschaften {\bf 21}, 787 (1933)
 \bibitem{london} F. London and H. London, Proc.Roy.Soc. {\bf A149}, 71 (1935).
 \bibitem{gauge} Eq. (1) is valid in the London gauge, $\vec{\nabla}\cdot \vec{A}=0$.
 \bibitem{meissner2} There was in fact considerable controversy in the literature initially on whether or not BCS theory can explain the Meissner effect in a gauge-independent way. 
 See G. Rickayzen, ``Theory of Superconductivity'', John Wiley $\&$ Sons, New York, 1965, and references therein.
 \bibitem{lenz}  J.E. Hirsch, Phys.Lett. {\bf A366}, 615 (2007).
  \bibitem{edwards} W. Farrell Edwards, Phys.Rev. Lett. {\bf 47}, 1863 (1981). 
  \bibitem{bostick} W. Bostick, Int. Jour. of Fusion Energy {\bf 3}, 47 (1985).
  \bibitem{witalis} E.A. Witalis, IEEE Trans. Plasma Sci. Vol. PS-14, p. 842 (1986).
  \bibitem{vahala} L. Vahala and G. Vahala, Physics Essays {\bf 4}, 223 (1991).
  \bibitem{heisenberg} W. Heisenberg, Z. fur Naturforschung {\bf 3a}, 65 (1948). London\cite{london2} showed this paper to be incorrect.
 \bibitem{london2} F. London, Phys.Rev. {74}, 562 (1948).
   \bibitem{putterman} S. Putterman, Phys.Lett. {\bf 89A}, 146 (1982).
 \bibitem{taylor} J.B. Taylor, Nature {\bf 299}, 681 (1982).
    \bibitem{lorentz}  J.E. Hirsch,  Phys.Lett. {\bf A 315}, 474 (2003).
 \bibitem{london35} H. London, Proc.Roy.Soc. London {\bf A152}, 650 (1035).
 \bibitem{nikulov} A.V. Nikulov, arXiv:cond-mat/0005121 (2000); Phys.Rev. B{\bf 64}, 012505 (2001).
  \bibitem{landau} A.A. Abrikosov, ``Fundamentals of the Theory of Metals'', North Holland, Amsterdam, 1988.
 \bibitem{gyro1} I.K. Kikoin and S.W. Gubar, J.Phys. USSR {\bf 3}, 333 (1940).
\bibitem{gyro2} R.H. Pry. A.L. Lathrop and W.V. Houston, Phys.Rev. {\bf 86}, 905 (1952).
\bibitem{gyro3} R. Doll, Zeitschrift fur Physik {\bf 153}, 207 (1958).
\bibitem{faber} T.E. Faber, Proc.Roy.Soc.  London {\bf A223}, 174 (1954) and references therein.
\bibitem{faber2} T. E. Faber and A. B. Pippard, Prog. Low Temp. Phys. {\bf VI}, 172 (1955). 
 \bibitem{lenz2} Furthermore the Faraday electric field generating by lowering the external magnetic field points in direction opposite to what is needed
 to account for the missing angular momentum.
 \bibitem{fogler} M. Fogler, private communication.
 \bibitem{anderson} P.W. Anderson, J. Phys. Chem. Solids {\bf 11}, 26 (1959).
    \bibitem{sm}  J.E. Hirsch,   arXiv:0710.0876  (2007), Europhys. Lett. {\bf 81},  67003 (2008).
 \bibitem{hole1} J.E. Hirsch and F. Marsiglio, Phys.Rev. {\bf B39}, 11515 (1989).
 \bibitem{holetheory} J.E. Hirsch,   Jour.Phys.Chem. Solids {\bf 67}, 21 (2006) and references therein.
  \bibitem{chargeexp} J.E. Hirsch, Phys.Rev.B {\bf 68}, 184502 (2003).
\bibitem{ionizing} J.E. Hirsch,  J.Phys. Cond. Matt.   {19},  125217 (2007).
 \bibitem{feynman} "The Feynman Lectures on Physics", R.P. Feynman, R.B. Leighton and M. Sands", Addison-Wesley,
Reading, 1964, Sect. 17-4.  
\bibitem{unpub} J.E. Hirsch, arXiv:0803.1198  (2008).
\bibitem{berger} J. Berger, Found. of Phys. Lett. {\bf 17}, 1572 (2004).
\bibitem{dipole}  J.E. Hirsch, Phys.Rev.B {\bf 42}, 4774 (1990).
 \bibitem{alfven}  W.A. Newcomb, Ann. of Phys. {\bf 3}, 347 (1958).
 \bibitem{fm}  J.E. Hirsch, Phys.Rev.{\bf B60}, 14787 (1999);  see also  arXiv:0709.1280 (2007).
  \bibitem{sh}   J.E. Hirsch, Phys.Rev.Lett.{\bf 83}, 1834 (1999).
    \bibitem{kuhn} T.S. Kuhn, ``The Structure of Scientific Revolutions'' (Univ. of Chicago Press, Chicago, 1970).
   \bibitem{lg} A. Lightman and O. Gingerich, Science {\bf 255}, 690 (1992).
 \bibitem{exp} J.E. Hirsch, Phys.Lett. {\bf A281},  44 (2001).  
 \bibitem{atom} J.E. Hirsch,  Phys.Lett. {\bf A 309}, 457 (2003).


 \end{references}
\end{document}